# High-performance $Ba_{1-x}K_xFe_2As_2$ superconducting tapes with grain texture engineered via a scalable and cost-effective fabrication

Shifa Liu[1,2], Chao Yao[1,*], He Huang[1], Chiheng Dong[1], Wenwen Guo[1,2], Zhe Cheng[1,2], Yanchang Zhu[1], Satoshi Awaji[3], Yanwei Ma[1,2,*]

[1]Key Laboratory of Applied Superconductivity, Institute of Electrical Engineering, Chinese Academy of Sciences, Beijing 100190, China

[2]University of Chinese Academy of Sciences, Beijing 100049, China

[3]High Field Laboratory for Superconducting Materials, Institute for Materials Research, Tohoku University, Sendai 980-8577, Japan

*Corresponding authors. E-mail: yaochao@mail.iee.ac.cn; ywma@mail.iee.ac.cn

**ABSTRACT**

Nowadays the development of high-field magnets strongly relies on the performance of superconducting materials. Iron-based superconductors exhibit high upper critical fields and low electromagnetic anisotropy, making them particularly attractive for high-field applications, especially in particle accelerator magnets, nuclear magnetic resonance spectrometers, medical magnetic resonance imaging systems and nuclear fusion reactors. Herein, through an industrially scalable and cost-effective manufacturing strategy, a practical level critical current density up to $1.1 \times 10^5$ A cm$^{-2}$ at 4.2 K in an external magnetic field of 10 T was achieved in Cu/Ag composite sheathed Ba$_{0.6}$K$_{0.4}$Fe$_2$As$_2$ superconducting tapes. The preparation strategy combines flat rolling to induce grain texture and a subsequent hot-isostatic-pressing densification. By varying the parameters of rolling, the degree of grain texture was engineered. It is found that the transport properties of the Ba$_{1-x}$K$_x$Fe$_2$As$_2$ tapes can be enhanced by applying a large amount of deformation during rolling, which can be attributed to the improved degree of c-axis texture. Microstructure characterizations on the highest-performance tape demonstrate that the Ba$_{1-x}$K$_x$Fe$_2$As$_2$ phase has a uniform element distribution and small grains with good connectivity. Grain boundary pinning is consequently enhanced as proved by large currents circulating through the sample even at 25 K. Our work proves that Cu/Ag composite sheathed Ba$_{1-x}$K$_x$Fe$_2$As$_2$ superconducting tapes can be a promising competitor for practical high-field applications in terms of the viable, scalable and cost-effective fabrication strategy applied and the high transport properties achieved in this work.

**INTRODUCTION**

Iron-based superconductors (IBSs) discovered in 2008 have attracted much attention due to their unique superconducting mechanism and potential value in practical applications [1-3]. To date, among several tens of reported IBSs [4-6], '122' iron-pnictide compounds such as $Sr_{1-x}K_xFe_2As_2$ (Sr122) and $Ba_{1-x}K_xFe_2As_2$ (Ba122) hold great promise for high-field magnet applications [7, 8], because they have moderately high critical transition temperatures $T_c$ (up to 38 K), high upper critical fields $H_{c2}$ (above 75 T) and low anisotropy $\gamma$ (<2) [8-11]. For such applications, superconducting wires and tapes with high in-field transport critical current densities ($J_c$) are crucial. Tremendous advances have been made in $J_c$ enhancement of iron-pnictide wires and tapes prepraed by a powder-in-tube (PIT) method over the past decade [4, 12, 13]. In order to obtain high $J_c$, besides high-quality precursor powder, there are two major issues that need to be resolved, namely the poor connectivity of Ba122 grains due to the microstructural defects such as pores and cracks which are extrinsically induced during the PIT process and intrinsic weak links at the high-angle grain boundaries between mismatched grains [14, 15]. Pores and cracks hinder the supercurrent flow inside superconducting polycrystals, while weak-link behavior results in an exponential decay of inter-grain $J_c$ as a function of grain boundary angle when it is larger than a so-called critical angle ($\theta_c$) [16]. Actually, Ba122 exhibits a much higher tolerance for mismatched grains than $REBa_2Cu_3O_7$ ($RE$=rare earth elements) cuprate superconductors [17-19]. The $\theta_c$ for cobalt-doped $BaFe_2As_2$ expitaxial films is about 9º, larger than the $\theta_c$ of 5º for $YBa_2Cu_3O_{7-\delta}$. Even for misorientation angles larger than $\theta_c$, cobalt-doped Ba122 films show much slower decay of $J_c$ than $YBa_2Cu_3O_{7-\delta}$. Measurements on Ba122 bulks reveal that a large $J_c$ results from small grains [20].

This attribute even has called into question whether grain texture is still crucial or not for high $J_c$-performance in iron-pnictide wires and tapes.

In '122' iron-pnictide tapes, the c-axis texture can be mechanically introduced by deforming round wires into thin tapes through flat rolling [21, 22]. In order to explicitly clarify the role of grain texture in $J_c$ improvement for Ba122 tapes, some extrinsic ingredients such as impurity phases, composition inhomogeneity, pores and cracks should be eliminated as much as possible. With uniaxial pressing densification process, a high transport $J_c$ above $10^5$ A cm$^{-2}$, which is a widely accepted threshold for practical applications, has been reported in c-axis textured Sr122 and Ba122 iron-pnictide tape samples [23-25]. However, strictly limited by the size of pressing heads, this manufacturing route is problematic to be scaled up for processing long length wires and tapes. Moreover, the widely used uniaxial press, either cold press or hot press, not only enhances the mass density, but also further improves grain texture simultaneously, making it complicated to investigate the contribution of mass densification and grain alignment to the $J_c$-performance seperately.

As an alternative to uniaxial pressing for densification, the hot isostatic pressing (HIP) process utilizing high-pressure argon gas as media can effectively reduce the pores and cracks in superconducting cores. The $J_c$ values of ~$10^4$ A cm$^{-2}$ at 4.2 K and 10 T were measured in HIP round wires with randomly oriented grains [26, 27]. The mass density of iron-pnictide cores of wires and tapes can be indicated with Vickers hardness. It is reported that the mass density of Ba122 cores in HIP processed wires can be significantly enhanced to nearly 100%, showing Vickers hardness even higher than that in uniaxially pressed tapes [24, 26-29]. Compared with uniaxial pressing, HIP technique is much more flexible and economic for large-scale producing, since long length wires and tapes can be processed in winding state in space-limited furnaces. However, it can hardly induce any grain alignment as done by rolling and uniaxial pressing, and further efforts over the past several years have failed to improve $J_c$ values of HIP-treated round wires to a practical level at a field of 10 T.

For Ba122 superconducting tapes, grain alignment introduced by rolling process can be well preserved after heat treatment by the HIP densification process [30]. Herein, flat rolling and a subsequent HIP process were combined to obtain c-axis texture and well-connected superconducting grains. Since mass densification and c-axis texture were achieved through two separate processes, the effect of grain alignment on transport $J_c$ can be studied based on highly dense Ba122 phase. The starting wire diameters for rolling were varied to induce different degrees of texture in the purpose of revealing the correlation between grain textures and transport $J_c$ in Ba122 tapes. Finally, as a result, $J_c$ was enhanced to $1.1 \times 10^5$ A cm$^{-2}$ at 4.2 K in 10 T, which confirms the indispensability of texture for high performance in Ba122 tapes.

**EXPERIMENTAL SECTION**

Ba122 tapes were prepared by the *ex situ* PIT method. The preparation of Ba122 precursor was detailed in the reference [30]. Figure 1(a), 1(b) and 1(c) illustrate the PIT and cold working deformation process. The precursor was ground into powder using an agate mortar, and sealed into a silver tube with an outer diameter of 8.0 mm and an inner diameter of 5.0 mm. These steps were done in an argon-filled glovebox to avoid contact with oxygen and moisture. The silver tube was first swaged into an outer diameter of 3.4 mm and then drawn into a wire with an outer diameter of 1.93 mm. The wire was cut into several pieces and each piece was sealed into a copper tube with an outer diameter of 4.0 mm and an inner diameter of 2.0 mm. The Cu/Ag composite wires were again deformed by swaging and drawing. The final diameters of these wires were 1.3, 1.5, 1.7 and 1.9 mm respectively. After that, these wires were rolled into tapes with thickness of 0.3 mm after 4 or 5 passes. Finally, these tapes were sintered for 1 h at 740°C by hot isostatic pressing at 150 MPa in argon atmosphere.

The field-dependent critical current measurements were performed at the High Field Laboratory for Superconducting Materials at Sendai by a standard four-probe method, with a criterion of 1 $\mu$Vcm$^{-1}$. It needs to be pointed out that the test was carried out under decreasing fields. For further analysis, several short pieces of these Ba122 tapes were chemically treated to remove the metal sheath to obtain the bare Ba122 cores or were embedded in conductive resin and mechanically ground and polished. The Vickers hardness testing was performed on the polished cross sections by using the Wilson 402MVD tester. The magnetic field and temperature dependent resistance of the Ba122 cores was measured by a Quantum Design physical property measurement system (PPMS). The magnetic property of the tape was characterized by SQUID-VSM on the magnetic property measurement system (MPMS). X-ray diffraction (XRD, Bruker D8 Advance) measurements were conducted on the surfaces of the Ba122 cores with Cu K$_{\alpha1}$ radiation. The scanning electron microscope (SEM, Zeiss SIGMA) observation was performed on the cross sections of the Ba122 cores by mechanically breaking. By using an EDAX Hikari camera equipped on the SEM, the grain texture was quantitatively represented by the Electron Backscatter Diffraction (EBSD) technique. A vibratory polishing was applied on the sample for EBSD analysis to eliminate the stress of the polished surface. The element distribution on the polished cross section was analyzed by electron probe micro-analyzer (EPMA, JXA-iSP100). The magneto-optical (MO) imaging was performed on a cryostat (Montana Instruments) by placing a Bi-substituted iron-garnet film on the top of the polished superconducting core. Magnetic fields perpendicular to the sample surface were provided by a homemade magnet.

**RESULTS**

Figure 1(d) shows the optical images of the four Ba122 superconducting tapes rolled from round wires with different diameters of 1.3, 1.5, 1.7 and 1.9 mm. The as-obtained tapes are labeled as tape-1.3-mm, tape-1.5-mm, tape-1.7-mm and tape-1.9-mm respectively. The typical cross-sectional images present a Ba122 superconducting core surrounded by an inner silver barrier and an outer copper sheath. In order to quantify the amount of deformation during rolling, a reduction ratio $r$ can be defined as $r=(R_0-d)/R_0$, where $R_0$ is the initial diameter and $d$ is the final tape thickness (in this work $d=0.3$ mm). The calculated $r$ values for tape-1.3-mm, tape-1.5-mm, tape-1.7-mm and tape-1.9-mm are 0.77, 0.80, 0.82 and 0.84 respectively. Vickers hardness of the Ba122 cores was measured on the cross sections of the tapes. The average values for these tapes are 202, 197, 203 and 212 respectively, which are comparable to that of the previously reported HIP samples [27, 30]. The high Vickers hardness usually indicates a high mass density of the superconducting cores.

In Figure 2(a), $J_c$ values of the four Ba122 tapes were plotted as a function of the applied magnetic fields. Among the four tapes, transport $J_c$ of tape-1.9-mm is the highest, reaching $1.1\times10^5$ A cm$^{-2}$ at 4.2 K and 10 T, which is a record value for Cu/Ag composite sheathed IBS tapes and surpasses the threshold for practical applications. It is sitll above $10^5$ A cm$^{-2}$ even at 12 T, and slightly decreases to 9.4 ×10$^4$ A cm$^{-2}$ at 14 T, showing very weak field dependence. For tape-1.7-mm, tape-1.5-mm and tape-1.3-mm, the $J_c$ values are 9.5×10$^4$, 9.2×10$^4$ and 5.8×10$^4$ A cm$^{-2}$ at 4.2 K in 10 T respectively. It is obvious that when the initial diameter of as-drawn wires becomes smaller, i.e. the reduction raio is lowered, the $J_c$ values of the tapes decrease. Figure 2(b) shows the resistance transition curves of the four tapes in the self-field. The inset in Figure 2(b) shows the resistivity of the four tapes as a function of temperatures in a range from 25 K to 250 K. It is worth noting that the resistance measurements were performed on the bare Ba122 cores, thus the result reflects the intrinsic property of the Ba122 phase in the tapes. For comparison, the resistivity of each tape is normalized by dividing by its resistivity value at 40 K. A distinctive feature of the resistance transition curves is that the four curves almost overlap with a $T_c$ difference within 0.1 K. Therefore, it can be inferred that the Ba122 polycrystals in these tapes have little difference in the element content and phase purity. The onset $T_c$ values of the four tapes are all around 38.2 K, indicating that high-quality Ba122 phase was obtained in these tapes.

The grain orientation and phase composition for the Ba122 tapes were investigated by X-ray diffraction (XRD), as shown in Figure 2(c). The pattern of the randomly orientated Ba122 precursor powder is also presented for comparison. Generally, for all the samples, Ba122 phase is the main phase, while in the tape samples, diffraction peaks of iron were also detected, which is ascribed to unreacted iron formed in HIP sintering. The intensity of all the XRD patterns was normalized by the intensity of the (103) peak of Ba122 phase. Compared to the precursor powder, the intensity of (00$l$) peaks in these tapes is strongly increased, which is a significant evidence of the presence of c-axis texture in these tapes. According to the Lotgering method [31], the degree of texture can be evaluated by an orientation factor $F=(\rho-\rho_0)/(1-\rho_0)$, where $\rho=\sum I(00l)/\sum I(hkl)$, $\rho_0=\sum I_0(00l)/\sum I_0(hkl)$, $I$ and $I_0$ are the intensities of each reflection peak ($hkl$) in the range of $2\theta$ from 10º to 60º for the tape samples and the randomly orientated precursor powder respectively. For tape-1.3-mm, tape-1.5-mm, tape-1.7-mm and tape-1.9-mm, the calculated $F$ values are 0.38, 0.42, 0.42 and 0.46 respectively. In Figure 2(d), the transport $J_c$ (4.2 K, 10 T) and Lotgering orientation factor $F$ are plotted as a function of the reduction ratio $r$. It clearly shows that the Ba122 tapes with a larger reduction ratio have a higher degree of texture and thus a higher transport $J_c$, and there is a positive and very strong correlation between grain texture and transport $J_c$. Considering the high Vickers hardness of around 200 for the superconducting cores in all the tapes, it can be inferred that high texture plays an important role for the realization of high $J_c$-performance.

In order to link the high transport properties in these tapes to their microstructure, detailed characterizations were carried out on the sample tape-1.9-mm, which possesses the highest $J_c$ value. First of all, the grain morphology the tape was viewed on the cross section by SEM, as shown in Figure 3(a, b, c). Figure 3(b) is a partially enlarged view of the area in the white dash-line frame. To describe the spatial relationship between the viewing angle and the tape, three principal axes are defined, the normal direction (ND), the rolling direction (RD) and the transverse direction (TD), as shown in the inset of Figure 3(b). The images clearly show the influence of rolling process on the grain alignment in the tapes. The maximum thickness of the core is about 60 μm with plate-like grains of about 5 μm in size. These plate-like grains are aligned along a direction approximately parallel to the tape surface due to the pressure component in the normal direction applied during the rolling process. Figure 3(c) show that the growth of Ba122 grains is suppressed near the interface between the silver barrier and the Ba122 core. Figure 3(e-i) show the element distribution maps of tape-1.9-mm on the polished cross section. The Secondary Electron Image (SEI) in Figure 3(d) shows a dense Ba122 phase embedded in the silver layer. A few defects were also observed. The white arrows indicate several small holes and the red oval indicates a barrium-riched partical as verified by the distribution map of barium. The partical in the green oval with a diameter of around 3 μm comes probably from the polishing suspension. Except for the barrium-riched dot, all the four elements (Ba, K, Fe and As) are uniformly distributed. The Ag mapping shows that near the interface between the Ba122 core and the Ag sheath, there is a 1~3 μm wide area where Ba-122 and Ag coexist, resulting in poor crystallinity for Ba-122 phase in this area, as indicated the white arrow in Figure 3(c). The EPMA results indicate that the Ba122 phase of tape-1.9-mm is dense with uniform element distribution, which is crucial for the realization of high transport $J_c$.

Despite the XRD and SEM results presenting strong evidences for the c-axis texture of the Ba122 grains in these tapes, quantitative and visualized characterizations through EBSD technique performed on the well-polished cross section of tape-1.9-mm can still provide valuable information. Figure 4(a) shows the inverse pole figure (IPF) map in (001) direction of a selected area viewed from the direction parallel to ND. The Ba122 grains are colored according to their grain orientations using a color cord given in the stereographic triangle. The dominate colors of the IPF map are red and reddish that correspond to c-axis texture. A quantitative statistics of the distribution of out-of-plane misorientation angles is shown in Figure 4(c). In this sample, the majority of the out-of-plane misorientation angles are distributed from 5° to 40°, showing a relatively low degree of texture than in hot-pressed tapes [25]. The inset in Figure 4(c) is the IPF of grain orientations.

Figure 4(b) presents Ba122 grains in the same location as figure 4(a) with gray scale ranging from black to white indicating the grain size. Grain boundaries in figure 4(b) are colored based on the grain misorientation angles, as shown in the legend below. The distribution of grain size is given in Figure 4(d) in two ways: number fraction and area fraction. The grain size ranges from 0.1 μm to 2.0 μm, which is smaller than in hot-pressed tapes [25]. Grains around 0.5 μm in size account for the largest proportion of the total area, and most of them are roughly c-axis textured. Besides, there are also a few relatively larger grains, about 1~2 μm in size. However, as Figure 4(a) and (b) show, lots of tiny grains (smaller than 0.2 μm in size) are radomly orientated with high-angle grain boundaries, which is definitely an impediment of the transport currents. This can be attributed to the relatively low sintering temperature and short holding time of the HIP process that may not be beneficial to the growth of grains.

MO imaging is a powerful tool that can visualize the distribution of magnetic flux. Through this method, one can directly observe the real-time distribution of magnetic flied in IBS superconductors and gain a wealth of information on homogeneity, grain connectivity and the intra- and intergranular $J_c$. The MO sample is shown in Figure 5(a). The sample dimension is 1.75 ×1.30 ×0.025 mm$^3$. Figure 5(b-f) show the MO images at Meissner state acquired by applying an external field on the zero-field cooled sample. At fields less than 125 Oe, the magnetic flux was excluded from the bulk due to the large shielding currents, except for a breakthrough at a crack produced by polishing. The flux begins to penetrate gradually into the whole sample at 314 Oe. Figure 5(c-f) clearly show the flux entering from the edges into the center in a dendritic path. However, flux is still precluded from the center region even at the maximum attainable field, indicating a large circulating current inside the Ba122 bulk. The remanent state obtained by decreasing the field from 1568 Oe to zero between 5 and 30 K are shown in Figure 5(e-j). A rooftop pattern of the magnetic flux was observed up to 30 K, indicting a uniform global current flow throughout the whole sample. It is noteworthy that the pattern of the trapped flux changes little below 20 K. Even at 25 K, the magnetic flux still cannot penetrate the entire sample, indicating that tape-1.9-mm has pretty good performance at high temperatures.

The resistance transition curves of the sample tape-1.9-mm in different magnetic fields were measured to evaluate the upper critical field $H_{c2}$, anisoropy $\gamma$ and the pinning potential $U_0$. Figure 6(a) shows the $R$-$T$ curves of different magnetic fields in directions of B//tape and B⊥tape, showing a transition broadening with increasing magnetic fields. The upper critical field $H_{c2}$ and the irreversibility field $H_{irr}$ are defined by 90% and 10% of the normal state resistivity, as shown in Figure 6(b). By using the Werthamer-Helfand-Hohenberg formula $H_{c2}(0)=-0.693T_c(dH_{c2}/dT)$ [32], with $dH_{c2}/dT$ at $T=T_c$, the $H_{c2}^{//}(0)$ and $H_{c2}^{\perp}(0)$ are estimated to be 240 T and 110 T, which are very high values even compared to (Ba, K)Fe$_2$As$_2$ single crystals [33]. The anisotropy is defined as $\gamma=H_{c2}^{//}/H_{c2}^{\perp}$ and varies from 2.0 to 3.6 at temperatures from 37.7 to 38.4 K, as plotted in the inset. The results are quite similar to those in (Ba, K)Fe$_2$As$_2$ single crystals, suggesting that Ba122 has small anisotropy as its intrinsic property. Furthermore, since the resistance in the low-resistance region is induced by the creep of vortices [34], the temperature dependences of the resistivity can be described as the equation $\rho(T,B)=\rho_0\exp[-U_0/k_BT]$, where $\rho_0$ is a parameter, $k_B$ is the Boltzmann's constant and $U_0$ is the flux-flow activation energy. By plotting ln$\rho$(T,B) as a function of 1/$T$, one can obtain $U_0/k_B$ from the slop of the linear part, as shown in Figure 6(c). The values of the activation energy $U_0/k_B$ range from 8500 K at 0.1 T to 3100 K at 9 T for the parallel fields and from 7500 K at 0.1 T to 2500 K at 9 T for the vertical fields, showing relatively weak field dependence compared to MgB$_2$ [35]. For comparison, $U_0$ values for Sr122 [36], YBCO [37], Bi2212 [38] and MgB$_2$ [35] are included in Figure 6(e). To summarize, these results indicate that our Ba122 tapes have very high intrinsic flux-pinning strength, high upper critical field $H_{c2}$, low anisotropy $\gamma$ and large activation energy $U_0/k_B$, which are the fundamental reasons for the high performance in high magnetic fields.

The flux pinning mechanism in tape-1.9-mm was analyzed base on the Dew-Hughes model [39]. First of all, the isothermal hysteresis loops of the tape was measured and the magnetic critical current density $J_c^{mag}$ was calculated based on the Bean model with the equation $J_c^{mag}=20\Delta M/w(1-w/3l)$ [40], where $\Delta M$ is the difference between the upper and lower branches of the hysteresis loops, $w$ and $l$ are the width and length of the sample ($l>w$), respectively. The irreversibility field $H_{irr}$ is determined by the linear part of the Kramer plot: $K_r(B)=J_c^{1/2}B^{1/4}$ at $K_r=0$, as shown in the inset of Figure 7. Figure 7 presents the normalized pinning force density $f=F_p/F_{p,\,max}$ as a function of $h=H/H_{irr}$. The data points can be fitted by the equation $f=Ah^p(1-h)^q$ with $p=0.47$, $q=1.87$. The fitting curve peaks at $h_0=0.20$, indicating that the pinning mechanism is more close to the grain boundary pinning when compared with our previous hot pressed tape whose $h_0$ locates at 0.22 [25, 27, 41]. This is in accordance with the fact that large amount of small grains caused by low temperature sintering exist in the high-performance tape. Since small grains are favorable for enhancing the pinning force, combing the results of EBSD analysis, one can conclude that these c-axis textured small grains ~0.5 μm are the desired ones and account for the high transport performance.

**DISCUSSION**

One of the major achievements of this work is that the transport $J_c$ exceeding $10^5$ A cm$^{-2}$ (4.2 K, 10 T) was achieved in Cu/Ag composite sheathed Ba122 tapes, which is currently a record value for low-cost Cu/Ag composite IBS tapes and wires. This value also exceeds a widely accepted threshold for practical applications, not to mention that a scalable fabrication route was applied. All of these make this work quite meaningful. Besides, relevant characterizations have shown that these tapes have pure Ba122 phase with high density and good uniformity. In this case, because the extrinsic factors were eliminated as much as possible, we emphasized the significant role of grain texture in achieving high transport $J_c$. Another distinct feature of our tapes is the small grain size (mostly ~ 0.5 μm). Based on the pinning mechanism analysis showing a peak at $h_0$=0.20 in the $f(h)$ curve, we put forward that the textured small grains in our tapes are the main reason for high $J_c$.

Despite the many favorable properties in the highest-$J_c$ tape-1.9-mm, several performance-hindering features were also found, suggesting that the $J_c$ values of HIP-sintered Ba122 tapes are still far from reaching its maximum. As revealed in both XRD and EBSD analysis, HIP-sintered Ba122 tapes have a relatively low degree of texture than the hot-pressed tapes with record $J_c$ values as high as $1.5\times10^5$ A cm$^{-2}$ (4.2 K, 10 T) [24]. This can be ascribed to the different ways in which high pressure was applied. For HIP-sintered Ba122 tapes, the high isostatic pressure provided by argon may not contribute much to further improve the grain texture. In order to enhance texture, more attention should be paid to the optimization of rolling process since it is the main means of introducing pre-texture. Increasing the wire diameter before rolling in this work is a proven ways to obtain higher pre-texture in Ba122 tapes. Reducing the final tape thickness may also produce similar benefits, but will face some difficulties in deformation uniformity.

**CONCLUSION**

In summary, Cu/Ag composite sheathed Ba122 iron-pnicted tapes with practical level transport $J_c$ were prepared through a scalable and cost-effective fabrication route that includes *ex situ* PIT method, cold rolling process and HIP technique. By tuning the rolling induced grain alingment, it is found that the transport $J_c$ of the tapes is closely and positively correlated to the degree of grain texture, proving that grain texture plays a critical role in improving $J_c$ for iron-based superdondcuting wires and tapes. Detailed characterizations show that the highest-$J_c$ sample tape-1.9-mm has pure Ba122 phase with high density and good uniformity, excellent grain connectivity, favorable performance at temperature up to 25 K, ultrahigh $H_{c2}$ up to 240 T, low anisotropy and high pinning potential. Quantitative analysis by means of EBSD reveals that c-axis textured small grains (~ 0.5 μm) account for the high transport $J_c$. On the other hand, several performance-hindering features were also found, such as relatively low degree of texture compared with hot-pressed tapes and randomly orientated tiny grains (less than 0.2 μm). Therefore, there is still room for the enhancement of the transport properties of Cu/Ag composite sheathed Ba122 tapes. This work proves the feasibility of large-scale low-cost preparation of high-performance IBS superconductors and will certainly promote the practical research of iron-based superconductors.

**ACKNOWLEDGMENTS**

This work is supported by the National Natural Science Foundation of China (Grant Nos. 51861135311, U1832213 and 51721005), the National Key R&D Program of China (Grant Nos. 2018YFA0704200 and 2017YFE0129500), the Strategic Priority Research Program of Chinese Academy of Sciences (Grant No. XDB25000000) and the Key Research Program of Frontier Sciences of Chinese Academy of Sciences (Grant No. QYZDJ-SSW-JSC026).

**Author Contributions**

Liu S designed the experimental plan, and performed sample preparation and most of the characterizations. Yao C and Dong C provided significant guidance on data analysis and co-authored this article. Huang H and Guo W measured the critical currents of the Ba122 tapes at 4.2 K under external magnetic fields. Cheng Z and Zhu Y contributed to the preparation of Ba122 precursor. Awaji S provided the high-field critical current test system and gave useful advice on the critical current measurements. Ma Y directed the project and also provided significant guidance on data analysis and article writing.

**Conflict of interest:** The authors declare that they have no conflict of interest.

**Captions**

**Figure 1.** (a, b, c) Illustrations of the fabrication process of Cu/Ag composite sheathed Ba122 tapes. (d) Optical images of typical cross sections of Cu/Ag composite sheathed Ba122 tapes. These tapes were flat rolled from round wires with diameters of 1.3, 1.5, 1.7 and 1.9 mm, respectively.

**Figure 2.** (a) Critical current density as a function of the applied magnetic field for tape-1.9-mm, tape-1.7-mm, tape-1.5-mm, tape-1.3-mm. (b) The temperature dependent resistivity of the tapes in the self-field. (c) XRD patterns of precursor powder and the cores of the tapes. The $Ba_{0.6}K_{0.4}Fe_2As_2$ diffraction peaks have been indexed. Solid rhombus indicates the diffraction peaks related to iron. (d) Transport $J_c$ (4.2 K, 10 T) and Lotgering orientation factor $F$ as a function of the reduction ratio $r$.

**Figure 3.** Typical SEM images showing the plate-like grain morphology of (a) the core of tape-1.9-mm, and the partial enlarged view (b) in the center as indicated by the white dash-line frame and (c) near the edge. The inset in (b) shows the three principal axes defined as ND, RD and TD. (d) The secondary electron image and (e-i) element distribution maps on the cross section of tape-1.9-mm.

**Figure 4.** Characterization of the grain structure of tape-1.9-mm viewed from the normal direction of the tape surfaces. (a) The IPF maps in [001] direction for tape-1.9-mm. (b) The grain size map on gray scale ranging from black to white on increasing grain size. The grain boundaries are colored based on the grain misorientation angles. (c) The distribution of out-of-plane misorientations in the selected area. The inset shows the the ND-IPF derived from IPF map in (a). (d) the distribution of the grain size of tape-1.9-mm presented in two ways: the number fraction and the area fraction.

**Figure 5.** (a) Optical micrograph of the MO sample with a dimension of 1.75 ×1.30×0.025 mm$^3$. (b-f) MO images showing flux penetration after zero-field-cooling the sample to 5 K and applying magnetic field. (g-l) The remanent state obtained by decreasing the field from 1568 Oe to zero between 5 and 30 K.

**Figure 6.** (a) The temperature-dependent resistivity of tape-1.9-mm in various magnetic fields for B// tape and B⊥tape. (b) $H_{c2}$ and $H_{irr}$ for both field directions. The inset presents the anisotropy $\gamma=H_{c2}^{//}/H_{c2}^{\perp}$ at different magnetic fields. (c) The Arrhenius plot to define the values of $U_0/k_B$. (d) The pinning potential $U_0/k_B$ as a function of magnetic fields.

**Figure 7.** Normalized pinning force density $f=F_p/F_{p,\,max}$ as a function of the reduced field $h=H/H_{irr}$. The inset shows the Kramer plot to define the $H_{irr}$. The black solid line is the fitting curve using the equation $f=Ah^p(1-h)^q$, which peaks at $h=0.20$.

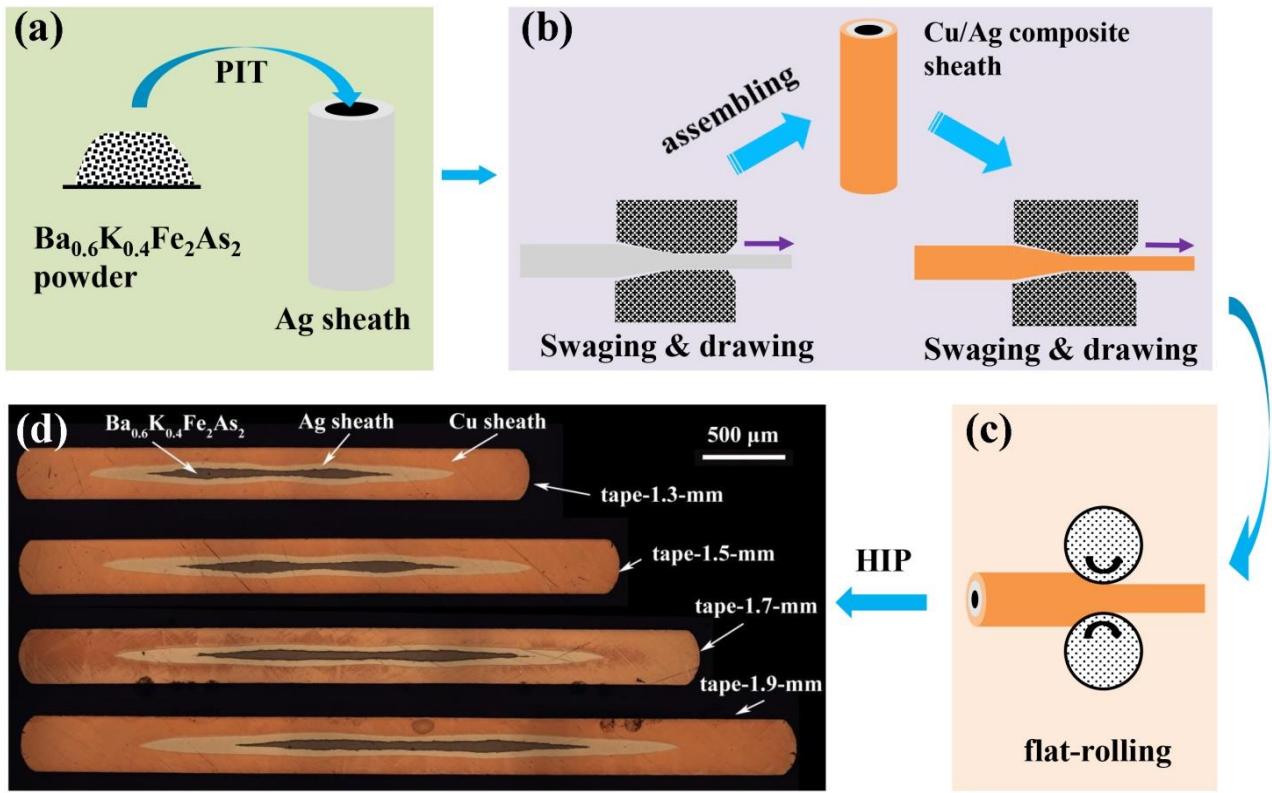

Figure 1

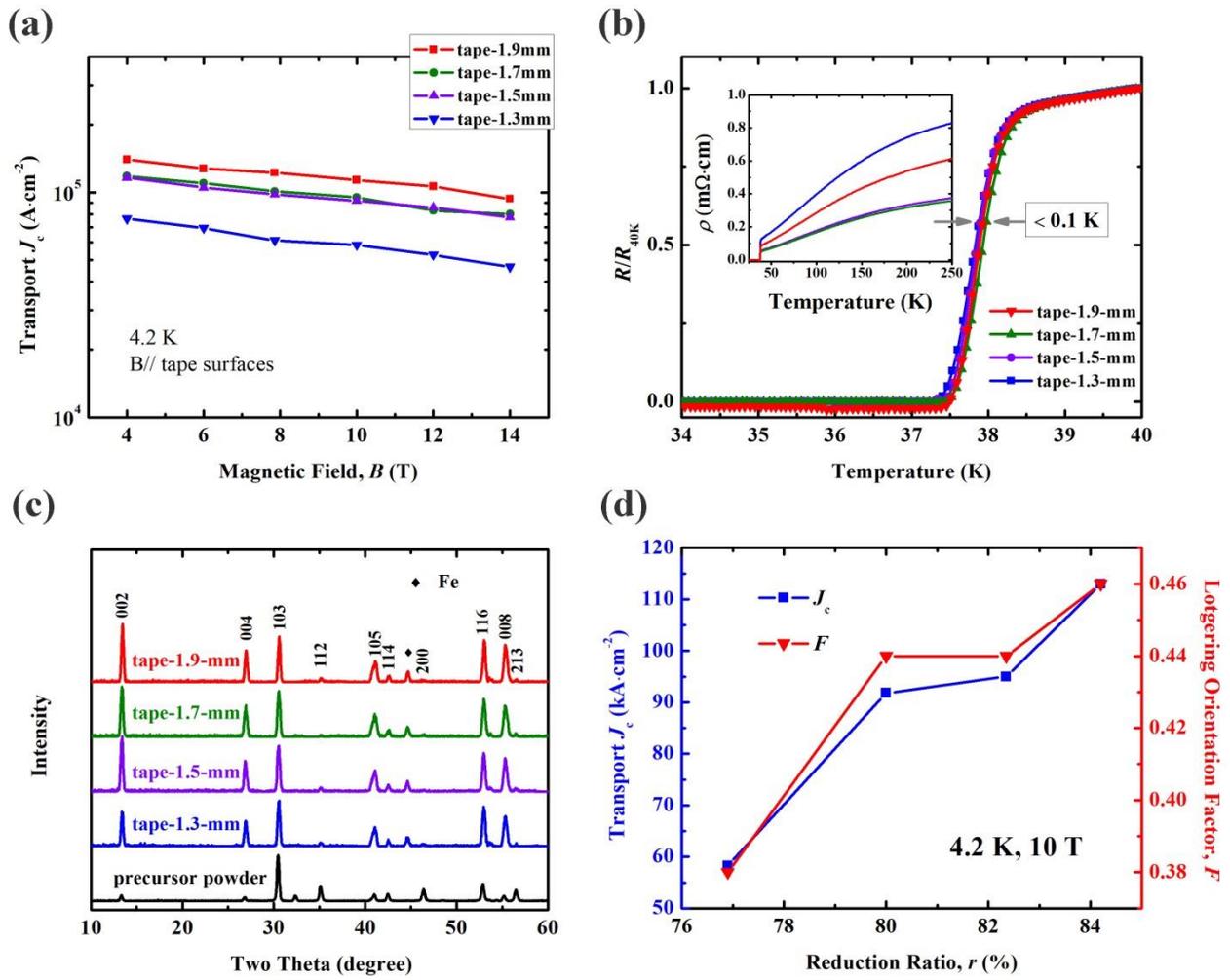

Figure 2

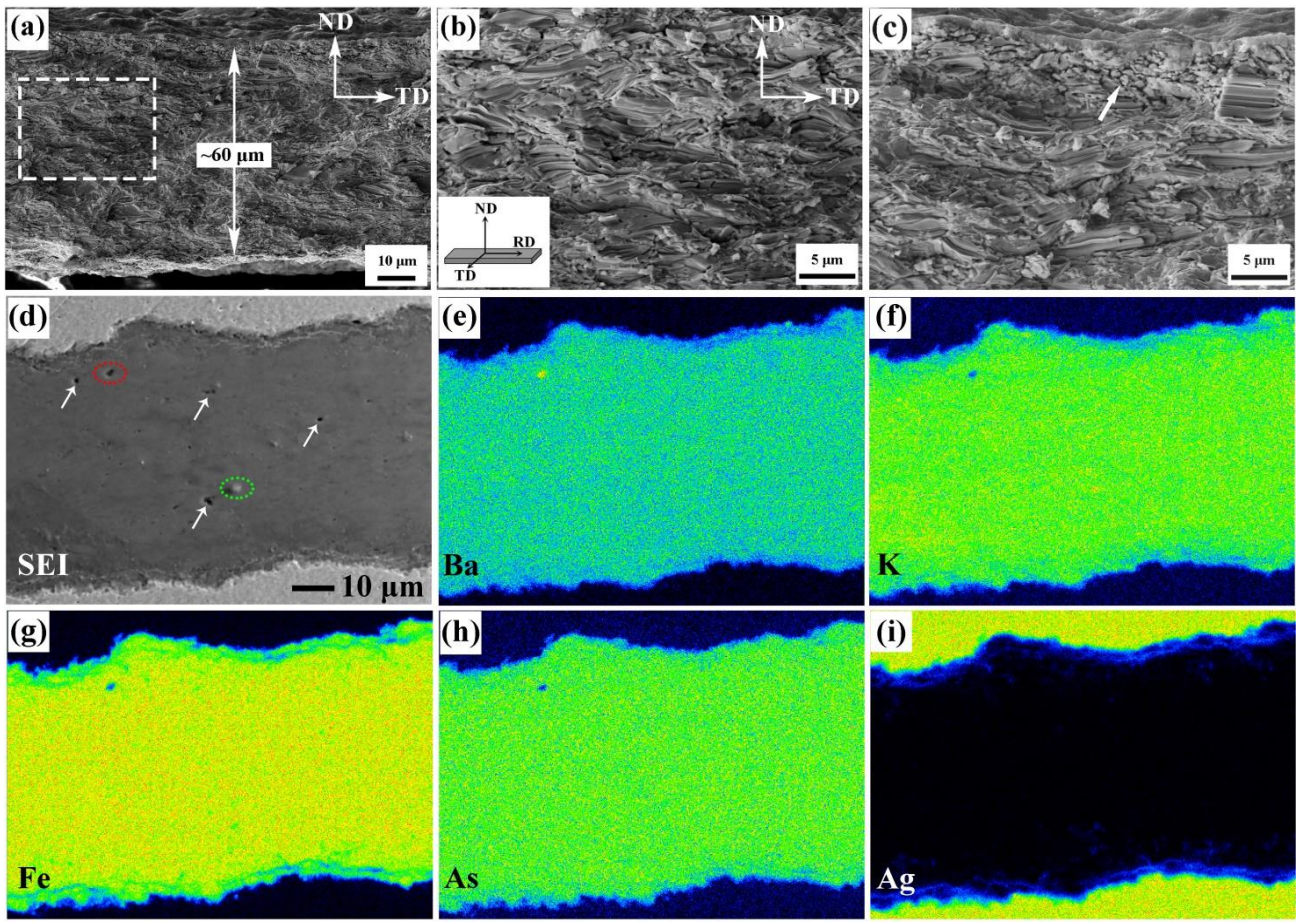

Figure 3

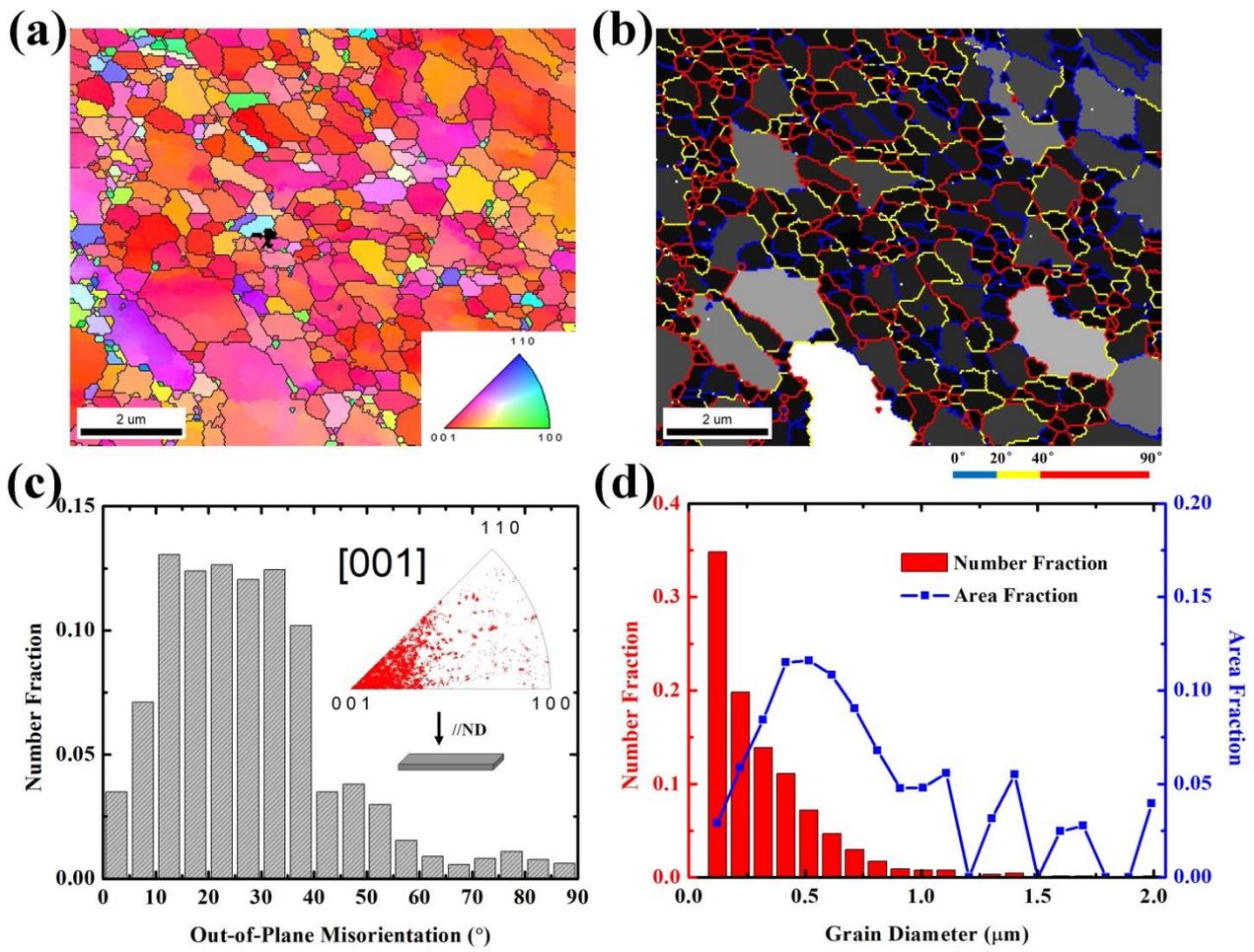

Figure 4

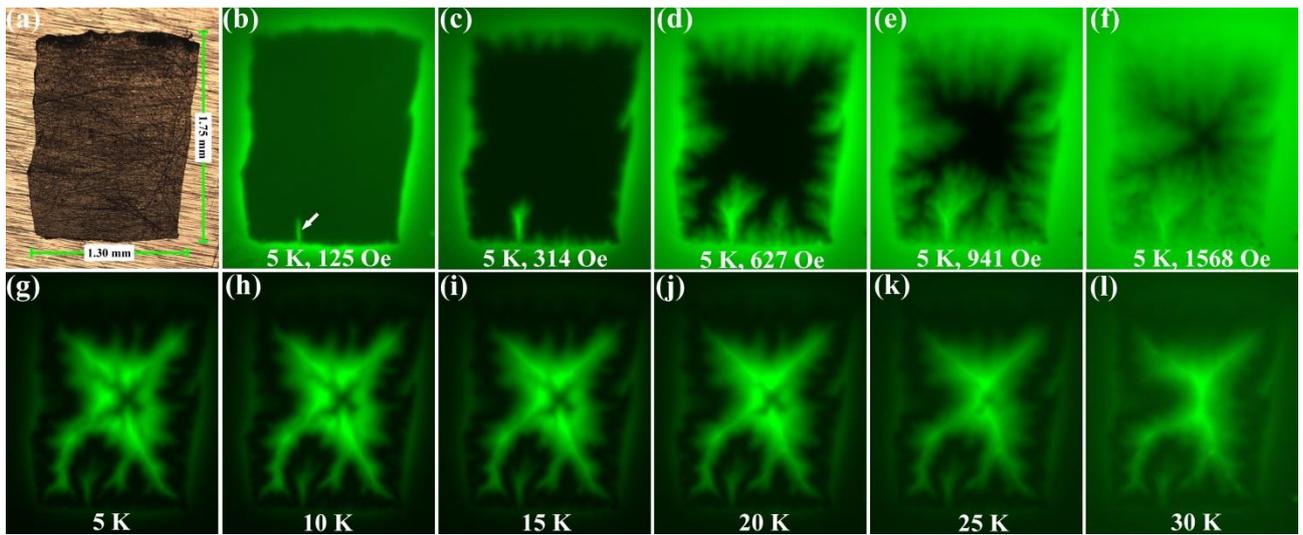

Figure 5

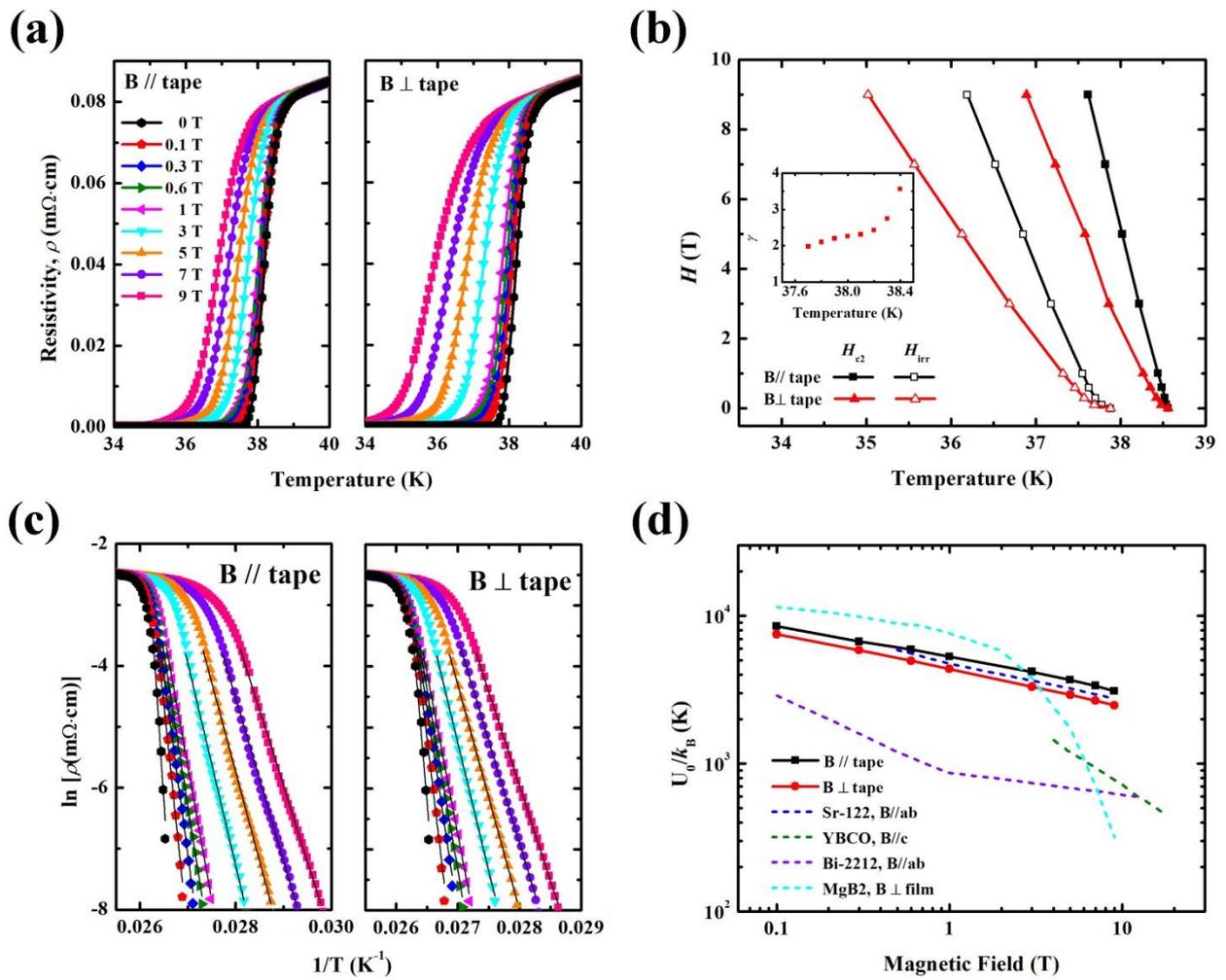

Figure 6

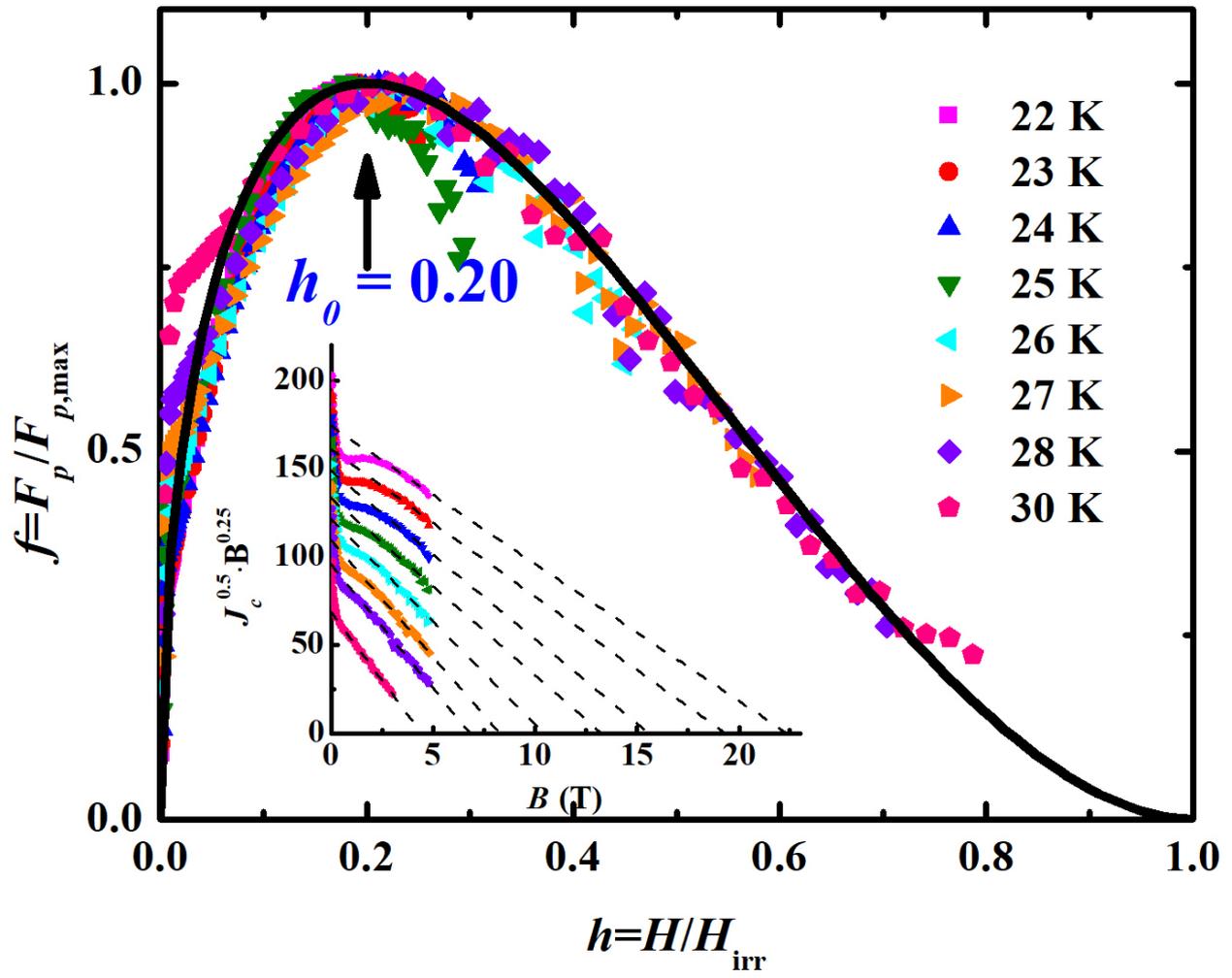

Figure 7